\documentclass[aps,prb,twocolumn, showpacs]{revtex4}

\usepackage{graphicx}
\usepackage{dcolumn}
\usepackage{amsmath}
\begin{document}

\title{Perturbative treatment of the multichannel interacting resonant level model in steady state non-equilibrium}
\author{L. Borda,$^{1,2}$ and A. Zawadowski$^2$} 
\affiliation{
             $^1$ Physikalisches Institut, Universit\"at Bonn,
                  Nu{\ss}allee 12 D-53115 Bonn, Germany\\ 
             $^2$ Department of Theoretical Physics,
                  Institute of Physics,
                  Budapest University of Technology and Economics,
                  Budapest, H-1521, Hungary} 
\date{\today}
\begin{abstract}
We consider the steady state non-equilibrium physics of the multichannel interacting resonant level model in the weak coupling regime. By using 
the scattering state method we show in agreement with the rate equations
that the negative differential conductance at large enough voltages is 
due to the renormalization of the hopping amplitude thus of the vertex.
\end{abstract}
\maketitle

Since the 
experimental
realization of quantum dots 
the electronic
transport in mesoscopic systems has attracted fastly growing
interest\cite{Qdots}. In these systems the short range Coulomb
interaction plays an important role and may result in strong correlations
between the electrons on the dot and the electronic leads. 

\begin{figure}[htb]
\includegraphics[width=0.75\columnwidth,clip]
{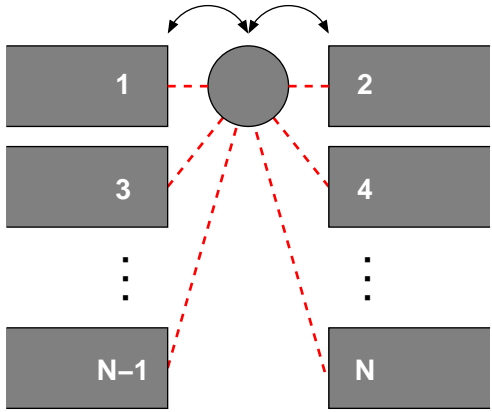}
\caption{Possible realization of the multichannel interacting resonant level model. A single level quantum dot is attached to several lead electrodes. Hopping
between the dot level and the neighboring sites of the leads $N=1,2$ are
present in addition to the short range Coulomb interaction acting
between the dot electron and the electrons on the neighboring sites of all $N$ electrodes.
The voltage is applied between leads 1 and 2.}
\label{fig:setup}
\end{figure}

In the following we consider a single level quantum dot with spinless
electrons thus only the interaction between the dot and the leads 
is taken into account.
The dot is attached to several lead electrodes thus hopping
between the dot level and the neighboring sites of the leads are
present in addition to the short range Coulomb interaction acting
between the dot electron and the electrons on the neighboring sites of
the leads as sketched in Fig.\ref{fig:setup}.

Without losing generality, we can assume that the hopping is
restricted to the leads 1 and 2 
while the Coulomb interaction acts to all of the $N$ leads. The model thus 
described by the Hamiltonian
\begin{eqnarray}
H&=&\sum\limits_{k,i=1\dots N}\varepsilon_{ki}c^\dagger_{ki}c^{\phantom{\dagger}}_{ki}+\varepsilon_d d^\dagger d\nonumber\\
&+&t_0\sum\limits_{k,i=1,2}d^\dagger
c^{\phantom{\dagger}}_{ki}+c^\dagger_{ki}d\nonumber\\
&+&{U\over2}\left(d^\dagger d-{1\over2}\right)\sum\limits_{k,i=1\dots N}\left(c^\dagger_{ki}c^{\phantom{\dagger}}_{ki}-{1\over2}\right)\;,
\label{eq:Hamiltonian}
\end{eqnarray}
where $c^\dagger_{ki}$ ($c^{\phantom{\dagger}}_{ki}$) creates (annihilates) an electron in the lead $i$ with momentum $k$, $d$ stands for the dot electron annihilation operator. $t_0$ is the hybridization amplitude between the dot and the leads 1 and 2, while $U$ is the strength of the Coulomb interaction acting between the dot and all the leads. The substraction of 1/2 in the last term is due to practical reasons: in this form the Hamiltonian obeys electron-hole symmetry if the local level is tuned to $\varepsilon_d=0$.

The equilibrium 
physics
of the model has been intensively studied
in the recent decades\cite{Vigman,Schlottmann,Varma,BVZ}. The
steady state non-equilibrium situation --beyond perturbation 
theory\cite{Doyon,BVZ}-- has been studied by exact and numerically
exact approaches
including Bethe Ansatz\cite{Pankaj}, conformal field theory and time dependent
density matrix renormalization group\cite{Bohr,BSS,Boulat}.

By applying these methods the I-V characteristics of the model has
been calculated where the external voltage applied between the
conduction electron leads labelled by 1 and 2 (left/right). Even if
the exact methods give correct answers it is hard to get insight into
the physical mechanisms and the different competing processes. The perturbative methods can provide help in the better understanding\cite{BVZ,Doyon}. Until now it was applied only to the relatively weak voltage bias case. Recently, at high bias a rather surprising phenomenon, namely a negative differential conductance has been suggested by exact methods\cite{Pankaj_thesis,BSS}. The aim of the present paper is to provide hints by perturbative methods how that negative differential conductance is established.

Following the earlier work the scattering state method combined with time dependent perturbation theory will be applied. That method is suitable for out of equilibrium situations as in the method any initial state can be considered. The renormalization is taken into account by a generalization of
Anderson's poor man's scaling including the corrections in next-to-leading logarithmic order and the renormalization of the initial and final states\cite{Solyom,BVZ}.

In the spinless model two competing phenomenon occur.
\begin{itemize}
\item[(i)] If the impurity level is occupied then the repulsive Coulomb interaction pushes away the electrons in the leads from the vicinity of the dot therefore creating more unoccupied states which helps the dot electron to tunnel to one of the leads. In this way the Coulomb interaction enhances the hopping rate.
\item[(ii)] When the occupation of the dot is changed then it has to be followed by the rearrangement of the electrons in the leads due to the Coulomb interaction. This rearrangement is similar to Anderson's orthogonality catastrophe takes a long time to be completed. This mechanism essentially reduces the hopping rate.
\end{itemize}

These two processes are competing in case of repulsive Coulomb interaction $U$. In perturbation theory the former one appears already in first order of $U$, while the latter one in second order only. Therefore a crossover is expected by increasing $U$.\cite{BVZ} In case of two leads that crossover occurs in the medium/strong coupling regime which is already outside of the validity region of the perturbative treatment. The crossover can be pushed 
down to the weak coupling regime by increasing the 
effect of the 
Coulomb repulsion by increasing the number of screening channels (leads) e.g. to $N=10$.

\begin{figure}[htb]
\includegraphics[width=0.75\columnwidth,clip]
                {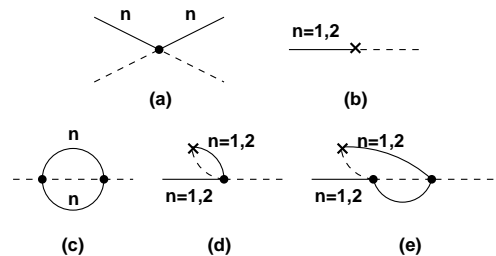}
\caption{The bare Coulomb vertex (a) and hybridization vertex (b). The solid lines represent the conduction electron propagators in different channels and the dashed line the electron on the dot. Diagram (c) is the self energy, (d) the vertex correction to the hopping in first order and (e) in second order. The flow of time agrees with the direction of the dotted line.}
\label{fig:diagrams}
\end{figure}

As in Ref.\onlinecite{BVZ} the self energy and the vertex corrections are given by
the skeleton diagrams in Fig.\ref{fig:diagrams} The solid lines represent the conduction electron propagators in different channels and the dashed line the electron on the dot. Diagrams (a) and (b) are the bare vertices (c) is the self energy, (d) the vertex correction to the hopping in first order and (e) in second order. The flow of time agrees with the direction of the dotted line.

The self energy and vertex corrections show different features concerning the role of the applied voltage. In the self energy diagrams the occupation of the dot is not changed thus the value of the applied voltage is irrelevant as only electron-hole pairs are created at one of the Fermi surfaces. In the vertex correction the occupation of the dot is changed by a hopping event thus the voltage $V$ is relevant and occurs as an infrared cutoff. 

The vertex correction in first order of $U$ (see Fig.\ref{fig:diagrams}(d)) 
is similar to a Hartree diagram thus it is independent of the incoming energy $\omega$ thus it is constant but depends on the voltage as $Ut\log[D/(eV+\omega_c)]$ where $\omega_c$ is the infrared cutoff arising from the width of the resonance\cite{footnote}. It is not necessarily the case for the correction of higher order. E.g. the correction depicted in
Fig.\ref{fig:diagrams}(e) gives
\begin{equation}
U^2t_0\log{D\over {eV}}\log{D \over{|\omega|+\omega_c}}
\label{eq:U2_in_eV}
\end{equation}
for $\omega\gg eV$. That may lead to difficulties to give the 
renormalization of the vertex for large
energies. However, in the transport only the electrons in the energy window $-eV/2 < \omega < eV/2$ play a role thus the appearance of such energy dependence is out of the focus of our interest.

\begin{figure}[htb]
\includegraphics[width=1.0\columnwidth,clip]
                {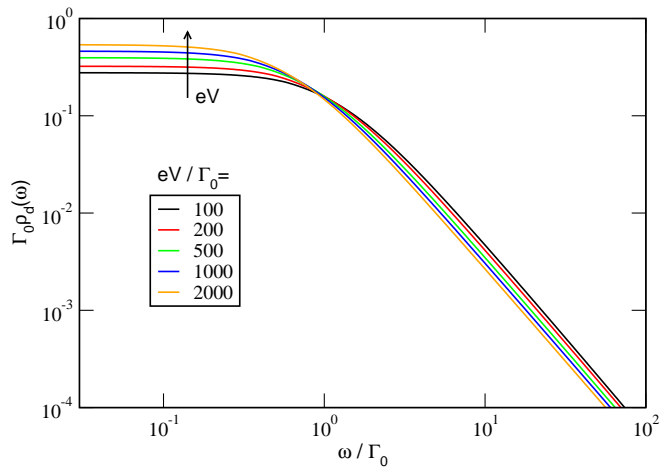}
\caption{The normalized impurity density of states for 
$N=10$ channels at $\varrho_0U=0.1$ and 
different values of the bias voltage. $\Gamma=2\pi\varrho_0t_0^2$.}
\label{fig:rho}
\end{figure}

The scaling equations provide the expression\cite{footnote}
\begin{equation}
t(\omega)=t_0\left({eV/2\over{D}}\right)^{-\varrho_0U}\left({|\omega|+\Gamma\over{D}}\right)^{N(\varrho_0U)^2/2}\;,
\label{eq:scaling}
\end{equation}
where $\varrho_0$ is the conduction electron density of states and it is assumed that the width of the resonance $\Gamma$ can be neglected in comparison with the applied voltage $eV$.

\begin{figure}[htb]
\includegraphics[width=1.0\columnwidth,clip]
                {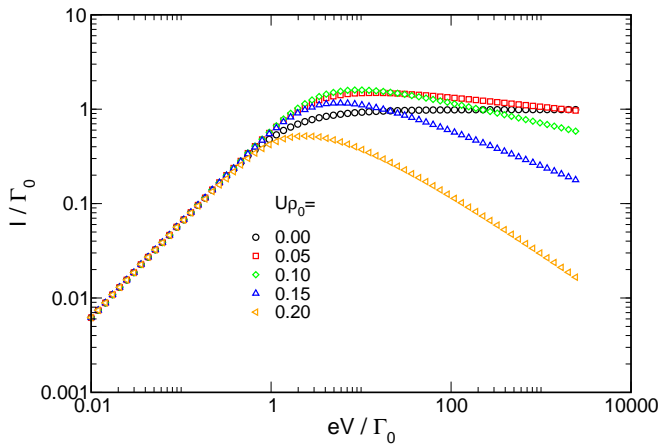}
\caption{$I-V$ characteristics with the clear evidence of the non-monotonic behavior for the $N=10$ channel interacting resonant level model for different values of the Coulomb repulsion.}
\label{fig:I_V}
\end{figure}

As in Ref.\onlinecite{BVZ} the rate equations 
are
used to determine the steady state current between the dot and one of the leads due to the voltage $eV$ applied on the leads (see Eqs. (28)-(31) of Ref.\onlinecite{BVZ}). In the scattering state method it is more feasible to calculate the scattering amplitude of an electron going from one of the leads to the other one. In the framework of the time ordered diagrams the first and last hopping must be picked up and the propagation of the electron on the dot can be described by the inverse life time $\Gamma(\omega)/2$ which contains the contribution to all orders in the hopping to the leads. In the initial state the impurity can either be occupied or unoccupied but the results are very similar. Thus that amplitude is
\begin{equation}
t(\omega){1\over{\omega+i\Gamma(\omega)/2}}t(\omega)
\label{eq:scatt_amplitude}
\end{equation} 
and in this way the current is
\begin{equation}
I(eV)=2\pi\varrho_0e\int\limits_{-eV/2}^{eV/2}t^4(\omega)\left|{1\over{\omega+i\Gamma(\omega)/2}}\right|^2 d\omega\;.
\label{eq:current}
\end{equation}
where the current flows in the energy window $-eV/2 < \omega < eV/2$ if the voltage bias applied symmetrically. $t(\omega)$ appearing in
\eqref{eq:current} is determined self consistently, i.e. its RG flow is terminated either by $\omega$, the voltage $eV$ or by $\Gamma(\omega)$ itself according to \eqref{eq:scaling}. 
Using the relation $2\pi\varrho_0t^2(\omega)=\Gamma(\omega)$ and 
$$
\varrho_d(\omega)={1\over\pi}{{\Gamma(\omega)/2}\over{\omega^2+(\Gamma(\omega)/2)^2}}
$$ 
we obtain 
\begin{equation}
I(eV)=2\pi e\int\limits_{-eV/2}^{eV/2}t^2(\omega)\varrho_d(\omega)\;.
\label{eq:current_2}
\end{equation}
Comparing Eqs.\eqref{eq:current} and \eqref{eq:current_2}
it is interesting to notice that a factor $t^2$ 
is now incorporated by the density of states $\varrho_d$. 
The result given by \eqref{eq:current_2} is just the current going to the dot from e.g. the left electrode. That holds only in the special case with L/R symmetry ($t_L=t_R$). It is worth mentioning that in this special case the results obtained by using rate equations and the scattering state method coincide. 

Considering the voltage dependence of the current two factors play role. According to Eq.(\ref{eq:scaling}) the voltage dependence of the hopping rate $t(\omega)$ is determined by the vertex correction which is reduced if the voltage is increased. On the other hand the renormalized density of states $\varrho_d$ is increasing with increasing voltage in the relevant energy window (see Fig \ref{fig:rho}). As the final result shows current reduction at large enough voltages, therefore the dominating process is the vertex renormalization (see Fig.\ref{fig:I_V}).

In conclusion, we have shown that remarkable physics of negative differential conductance in the interacting resonant level model previously studied by exact methods can be understood on the ground of perturbation theory. Such a treatment of the problem --though it is restricted to the weak coupling regime of the model-- permits us to gain insight into the mechanisms leading to negative differential conductance at large voltages. Our conclusion is that the underlying mechanism is the vertex renormalization, therefore, the
mechanism leading to a negative differential conductance 
takes its origin in 
the equilibrium dynamics of the model. As the interacting resonant level model has become a benchmark model for different theoretical methods dealing with non-equilibrium physics of quantum impurity problems, it would be nice
to check whether the more precise RG methods\cite{Rosch,Schoeller,Kehrein,Anders,Meden}  
provide conclusion similar to ours.

---{\em Acknowledgments}.
We are grateful to N. Andrei, P. W\"olfle, P. Schmitteckert, H. Schoeller, S. Andergassen, C. Karrasch and V. Meden for stimulating discussions. This research is supported in part by the Alexander von Humboldt Foundation (L.B.).


\begin{thebibliography}{99}
\bibitem{footnote} In Eq.(27) of Ref.\onlinecite{BVZ} the $\omega$-dependence
of the term linear in $U$ is incorrect.
\end{thebibliography}
\end{document}